\documentclass[aps,prl,reprint,superscriptaddress]{revtex4-2}
\usepackage{epsfig}
\usepackage{graphicx}
\usepackage[caption=false]{subfig}
\usepackage{bm}
\usepackage{color}
\usepackage{float}
\usepackage{mathtools}
\usepackage{amsmath}
\usepackage{comment}
\usepackage{textcomp}

\begin{document}

	\title{Nuclear Magnetic Resonance Investigation of Superconducting and Normal State Nb$_3$Sn}
	
	\author{Gan Zhai}
	\altaffiliation{ganzhai2025@u.northwestern.edu}
	\affiliation{Department of Physics and Astronomy, Northwestern University, Evanston IL 60208, USA}
	\author{William P. Halperin}
	\altaffiliation{w-halperin@northwestern.edu}
	\affiliation{Department of Physics and Astronomy, Northwestern University, Evanston IL 60208, USA}
	\author{Arneil P. Reyes}
	\affiliation{National High Magnetic Field Laboratory, Tallahassee FL 32310, USA}
	\author{Sam Posen}
	\affiliation{Fermi National Accelerator Laboratory, Batavia IL 60510, USA}
	\author{Zuhawn Sung}
	\affiliation{Fermi National Accelerator Laboratory, Batavia IL 60510, USA}
	\author{Chiara Tarantini}
	\affiliation{National High Magnetic Field Laboratory, Tallahassee FL 32310, USA}
	\author{Michael D. Brown}
	\affiliation{Bruker OST, Carteret NJ 07008, USA}
	\author{David C. Larbalestier}
	\affiliation{National High Magnetic Field Laboratory, Tallahassee FL 32310, USA}
	
	\date{\today}
	
	\begin{abstract}
	The superconductor Nb$_3$Sn has a high critical temperature and high critical field, widely used for high-field superconducting magnets. In this work we investigate its microscopic electronic structure with $^{93}$Nb nuclear magnetic resonance (NMR). The high-quality Nb$_3$Sn powder sample was studied in both 3.2\,T and 7\,T magnetic fields in the temperature range from 1.5\,K to 300\,K. From  measurement of the spectrum and its theoretical analysis, we find evidence for anisotropy despite its cubic crystal structure. This anisotropy is manifest in alignment of powder grains under certain temperature and field cycling conditions. The Knight shift and spin-lattice relaxation rate, $T_1^{-1}$, were measured in the normal state. Additionally, $T_1^{-1}$ was measured in the superconducting state and compared with  BCS theory revealing a weak field dependence, with an energy gap $\Delta(0)=2.0\pm0.08\,k_B T_c$ at 3.2\,T and $\Delta(0)=1.73\pm0.08\,k_B T_c$ at 7\,T, indicating suppression of the order parameter by magnetic field.
	\end{abstract}

	\maketitle
	
	\section{Introduction}
	Nb$_3$Sn is widely used for producing superconducting magnet wires. Its high critical temperature $T_c (\sim18\,$K) and high critical field $H_{c2} (\sim30\,$T)~\cite{Zho.11} make it possible to generate very high magnetic fields at liquid helium temperatures, such for NMR high resolution magnets operating in the 1 GHz range and hybrid magnets at the National High Magnetic Field Laboratory reaching 45 T. Significant amounts of Nb$_3$Sn,  $\sim 875$ tons, have been used for the ITER tokamak magnets~\cite{Dev.14}.  
Recent research shows that this material can also be used for superconducting radio frequency (SRF) cavities~\cite{Pos.21}.
	
	Nuclear magnetic resonance (NMR) has traditionally played a key role in the investigation of superconducting materials both in normal and superconducting states, in low and high magnetic fields and at low temperatures~\cite{Mac.76}. 
The electronic coupling to the nuclear spin provides direct information from NMR about the density of electronic states, the amplitude of the superconducting order parameter, and superconducting vortex structure~\cite{Mit.01}.
However, despite  important applications for Nb$_3$Sn, the electronic structure properties of this material have not been extensively studied with NMR. An early effort using $^{93}$Nb NMR in 1973~\cite{Fra.73} was incomplete, notably in the superconducting state.  We have overcome those earlier limitations and present our work here.
	
	\section{Sample preparation and characterization}
\begin{figure}[htbp]
		\includegraphics[width=0.5\linewidth]{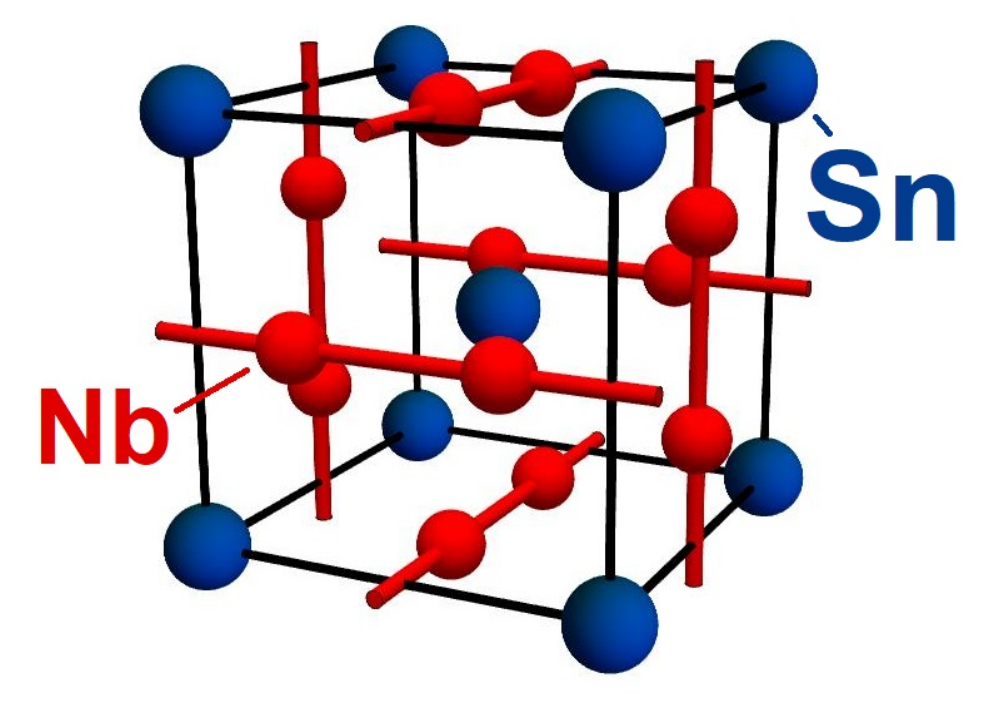}
		\caption{\label{Fig1} Crystal structure of Nb$_3$Sn.}
	\end{figure}

	The high-quality powder sample, 20 mg, was produced in the Applied Superconductivity Center of the National High Magnetic Field Laboratory following a process of high-energy ball milling, initial cold isostatic press densification, and final hot isostatic press reaction and densification at 1800\,\textcelsius~\cite{Zho.11}. Our sample was made from 27\% Sn and 73\% Nb expressed in atomic percent; therefore, it is slightly Sn-rich (T$_c\,=\,17.2\,$K, H$_{c2}\,=\,28.5\,$T at 1.2\,K). According to X-ray characterization (XRD) it has a typical cubic A15 crystal structure, shown in Fig.~\ref{Fig1}.  For this composition and processing there is no evidence of tetragonal structure. The Sn atoms occupy the corners and the body-centered positions of a unit cell, while the Nb atoms lie on three orthogonal chains on each face of the cube.

	\section{Experiment}
	The NMR experiments were performed at Northwestern University in a magnetic field of 14\,T, and at the National High Magnetic Field Laboratory with a magnetic field of 3.2\,T and 7\,T.
	
	Given that the Nb nucleus has a spin I=9/2, there are strong quadrupolar components to a very broad spectrum. For a random powder, this results in a very short signal decay time, less than the dead time of the spectrometer. Therefore, instead of a single free induction decay, we used a Hahn echo ($90^\circ-180^\circ$ tip angle technique) for all  NMR measurements. We swept the frequency over a wide range in order to capture the full spectrum.
	
	The superconducting transition temperature $T_c$ was measured to be $14.5\pm$0.2\,K at 7\,T and $16\pm$0.2\,K at 3.2\,T by observing the change of the tuning frequency of the NMR coil at the transition temperature. The temperature dependence of the  susceptibility was measured with a SQUID demonstrating absence of paramagnetic impurities.
	
	\section{Results}
	\subsection{Spectrum}
	
	
	\begin{figure}[htbp]
		\includegraphics[width=\linewidth]{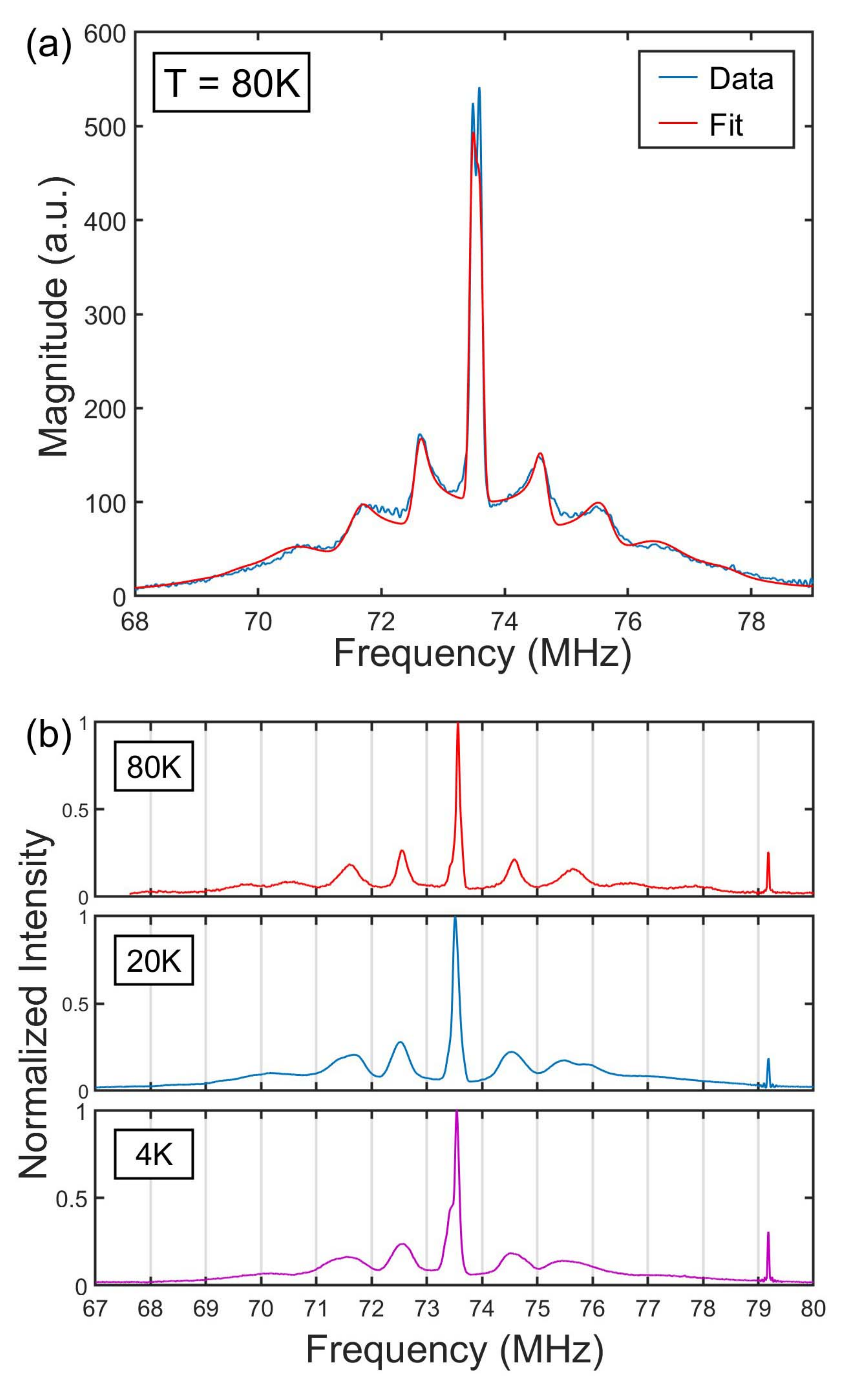}
		\caption{\label{Fig2} Nb$_3$Sn Spectra at 7\,T. (a) $^{93}$Nb spectrum at 80\,K field cooled. The red line is a fit with the theory, Ref.~\citealp{Jon.63}. (b) $^{93}$Nb spectra at 4\,K, 20\,K and 80\,K after cooling to 1.5\, K in low field and then ramping to 7\,T. The small peak on the right side is the $^{63}$Cu signal from the NMR coil. The quadrupolar broadening of the satellites is noticeably decreased, indicating an alignment effect in the powder sample. The dome-like background  in (a) results from a significant overlap of the satellites in the random powder.}
	\end{figure}

	The expected nine component spectrum of $^{93}$Nb was measured at 80\,K in a field of 7\,T, as shown in Fig.~\ref{Fig2}(a) for what we have determined from spectrum analysis is a randomly oriented powder. This can be compared with spectra at different temperatures from the oriented sample  shown in Fig.~\ref{Fig2}(b). The difference between the two cases (a) and (b) can be attributed to the field-temperature cycle producing preferential magnetic alignment. The shape of the spectra was analyzed using the theory given by W. H. Jones~\cite{Jon.63}. The line shape depends on an anisotropic Knight shift and the nuclear quadrupole coupling to the local electric field gradient, principally from near neighbors. It depends on the relative orientation of the crystal axes with respect to the magnetic field. We speculate, that ball milling Nb$_3$Sn fractures  the bulk polycrystalline material at crystal grain boundaries since the surfeit of Sn segregates at the grain boundary~\cite{Zho.11}. Alignment results from magnetic anisotropy which we discuss next. 	
	The anisotropic Knight shift can be expressed quantitatively by separating the Knight shift into isotropic and axial components, $K_{iso}$ and $K_{ax}$. We can then determine the central transition frequency $\nu_0=\nu_R (1+K_{iso})$ and its anisotropic factor $a=K_{ax}/(1+K_{iso})$.
	
	The nuclear quadrupole interaction depends on the nuclear quadrupole moment Q and the local electric field gradient {\bf q} (EFG) . The effect of the quadrupole interaction on the NMR frequency is expressed in terms of the quadrupole frequency, defined as $\nu_Q=3e^2qQ/2I(2I-1)h$ and is a measure of the separation in frequencies of the spectral components.
	
	The resonance frequency of a nucleus with spin $I\neq1/2$ in a single crystal, correct to  second order in perturbation theory, is:
	
	\begin{equation}
	\begin{split}
	\nu(m \leftrightarrow m-1)=&\nu_0+(3\mu^2-1)[a+1/2 \nu_Q (m-1/2)]\\
	&+1/32 ((\nu_Q^2)/\nu_0 )(1-\mu^2 )\\
	&\times\{[102m(m-1)-18I(I+1)+39]\mu^2\\
	&-[6m(m-1)-2I(I+1)+3]\}
	\end{split}
	\label{SpectrumFit}
	\end{equation}
	
	In this expression, $\mu \equiv \mathrm{cos}\,\theta$, where $\theta$ is the angle between the principal axis of the crystal and the external magnetic field. In a randomly oriented powder sample, the direction of the principal axis is equally distributed in space, producing a uniform distribution of $\mu$ from -1 to 1. With this distribution, we can simulate the spectrum by convolving the orientational frequency distribution function with a shape function, taken as a Gaussian distribution with a standard deviation $\Delta\nu_0$~\cite{Fra.73}. Note that $\Delta\nu_0$ includes both homogeneous and inhomogeneous broadening. The homogeneous component is determined by the intrinsic $T_2$, while the inhomogeneous component is dominated by the orientation distribution of the EFG. Hence in the simulation, $\Delta\nu_0$ is larger for outer satellites, with a ratio of 1:2:4:8:16, extending from the central transition to the outermost satellite.
	
	The fit to the spectrum in Fig.~\ref{Fig2}(a), shown in red, gives $K_{ax} = 0.5\%$, which represents the local anisotropy in the near vicinity of the Nb nucleus. In further experiments, we found that certain temperature and field cycling, ramping up the field at low temperature 1.5\,K, resulted in some degree of alignment. This is reflected in the spectra shown in Fig.~\ref{Fig2}(b), clearly distinct from that in Fig.~\ref{Fig2}(a). The three spectra in Fig.~\ref{Fig2}(b) deviate from a typical random powder spectrum, which has a dome-like background corresponding to a broad distribution of orientations of  electric field gradients. The line shape of the satellites are consistent with the distribution of $\mu$ to be enhanced at either the limits $\mu=0$ or $\mu=1$, indicating respectively that the powder grains are more, or less, aligned with the external field. This argument requires that a significant fraction of the powder grains have a high degree of crystallinity, as well as intrinsic anisotropy in the susceptibility.  Fig.~\ref{Fig2}(b) shows that when ramping up the temperature, the line shape is preserved from 4\,K to 80\,K, meaning that this ordering is stable and not related to the superconducting state. A detailed study of the alignment phenomenon will be undertaken in future work.

	\subsection{Normal state measurements}
	
	
	\begin{figure}[htbp]
		\includegraphics[width=\linewidth]{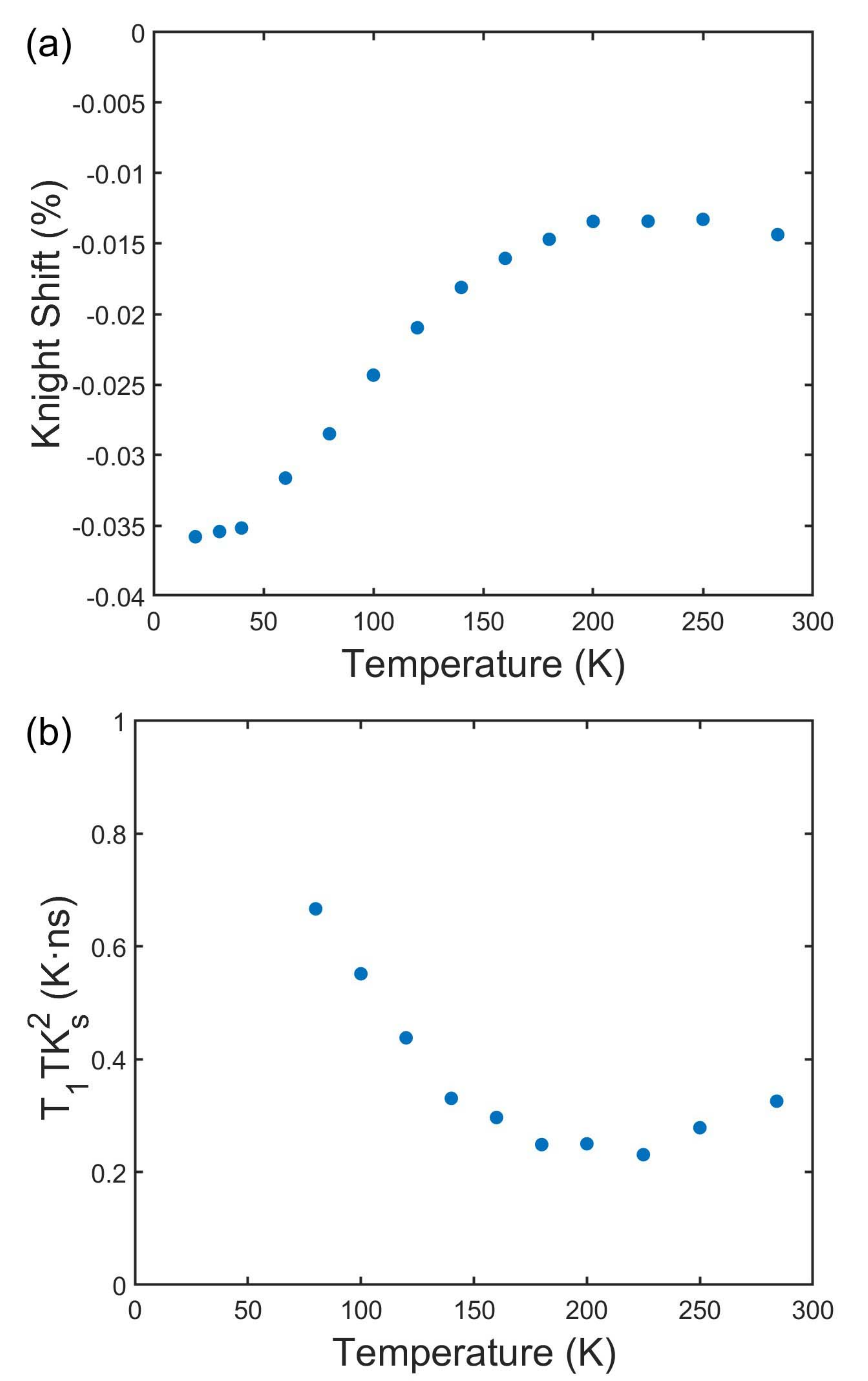}
		\caption{\label{Fig3} Temperature dependence of the $^{93}$Nb Knight shift (a) and Korringa relation (b) in the normal state at 14\,T.}
	\end{figure}

	The total frequency shift, $K$, in the normal state was measured at the central transition from room temperature to 20\,K. $K$ has two components, the spin component $K_s$, also known as the Knight shift, and the orbital and diamagnetic component $K_o$, given in Eq.~\ref{K}. The latter, $K_o$, is temperature independent and can be obtained from the zero temperature limit of the total shift K, since the spin component $K_s$ is fully eliminated due to the formation of Cooper pairs in the superconducting state. In this case we have $K_o=0.8835\%$ relative to the Larmor frequency. Then the Knight Shift $K_s$ can be extracted from the total shift and plotted in Fig.~\ref{Fig3}(a).
	
	\begin{equation}
		K = K_s + K_{o}
		\label{K}
	\end{equation}
	
	The small temperature dependence of the Knight shift can be attributed to core polarization of the Nb atomic orbitals. The decrease of the Knight shift with decreasing temperature indicates that the core polarization contribution to the shift is enhanced at lower temperatures. The dependence of the Knight shift and the spin lattice relaxation rate $T_1^{-1}$ are given in Eq.~\ref{KvsD} and Eq.~\ref{T1vsD}, the latter is known as the Korringa relation,  expressed in terms of the density of states, $\rho(E_F)$, at the Fermi surface compared to the free electron case, $\rho_0(E_F)$. 
	
	\begin{equation}
		K_s=\frac{8\pi}{3} \mu_0 {\mu_B}^2 \langle \lvert u_k (0) ^2 \rvert \rangle _{E_F} \rho(E_F)
		\label{KvsD}
	\end{equation}
	
	\begin{equation}
		T_1 T K_s^2 = \frac{\hbar}{4\pi k_B}\,\frac{\gamma_e^2}{\gamma_n^2} \left[ \frac{\chi_e^s}{\chi_0^s}\,\frac{\rho_0(E_F)}{\rho(E_F)} \right]^2
		\label{T1vsD}
	\end{equation}
	
	The spin-lattice relaxation rate was measured at the central transition, and the Korringa relation is plotted in Fig.~\ref{Fig3}(b). The measurements were performed with a saturation recovery pulse sequence ($90^\circ-90^\circ-180^\circ$), using a Hahn echo for detection as was discussed above. The $T_1$ was obtained from a stretched exponential fit $M=M_0 [1-\mathrm{exp}(-t/T_1 )^\beta]$ for the recovery of the NMR signal $M$ with $\beta\approx0.8$.
	
	\subsection{Superconducting state measurements}
	
	
	\begin{figure}[htbp]
		\includegraphics[width=\linewidth]{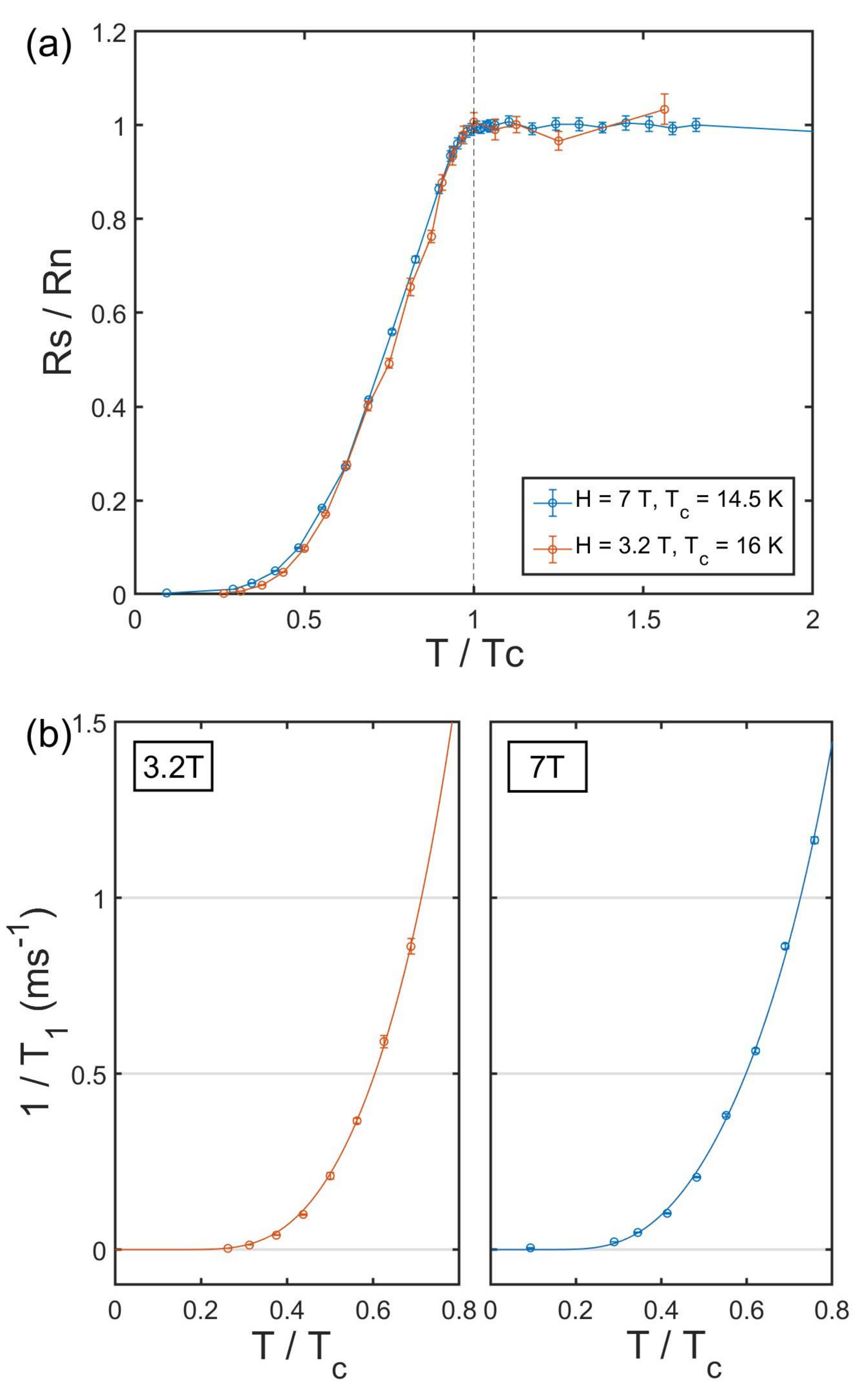}
		\caption{\label{Fig4} $^{93}$Nb spin-lattice relaxation rate $T_1^{-1}$ in the superconducting state of the Nb$_3$Sn powder. (a) Relaxation rate ratio vs. Temperature. The dots of two colors are data with error bars at 3.2\,T and 7\,T, shown separately). The lines are guides to the eye. (b) BCS fit for the temperature dependence of the relaxation rates at 3.2\,T and 7\,T, shown separately.  The fitting results are shown by curves giving $\Delta(0)=2.0\pm0.08\,k_B T_c$ at 3.2\,T and $\Delta(0)=1.73\pm0.08\,k_B T_c$ at 7\,T.}
	\end{figure}

	The spin lattice relaxation rate $R_s=T_1^{-1}$  was measured at the central transition in the superconducting state for both 3.2\,T and 7\,T, Fig.~\ref{Fig4}(a). It is normalized to the relaxation rate $R_n$ at $T_c$. To better reflect the normal state behavior of $T_1^{-1}$, $R_n$ is defined in the normal state as the average of $(T/T_c)T_1^{-1}$ and in the superconducting state set equal to its value at $T_c$. The ratio $R_s/R_n$ does not decrease immediately below $T_c$. This can be attributed to quasiparticle coherence. There is no evidence for a Hebel-Slichter peak\cite{Mac.76} for either 3.2\,T and 7\,T.  The temperature dependence of $R_s$ is fitted to the BCS theory, as shown in Fig.~\ref{Fig4}(b). The relaxation rate was analyzed in the low temperature limit, Eq.~\ref{T1vsDelta}, with the BCS energy gap $\Delta(T)$ phenomenologically expressed in Eq.~\ref{DeltavsT}~\cite{Hal.90}:
	
	\begin{equation}
	T_1^{-1} \propto \mathrm{exp}(-\Delta(T)/k_BT)
	\label{T1vsDelta}
	\end{equation}
	
	\begin{equation}
	\Delta(T) = \Delta (0) \mathrm{tanh} \{ \frac{\pi k_B T_c}{\Delta(0)} [\frac{2}{3} (\frac{T_c}{T}-1) \frac{\Delta C}{C}]^{1/2} \}
	\label{DeltavsT}
	\end{equation}
	
	In the above expression we take, $\Delta C/C=1.426$ as given by BCS weak-coupling  theory. Then the energy gap $\Delta(0)$ can be inferred by fitting the data to the expression above in the low-temperature limit. From Fig.~\ref{Fig4}(b), we get $\Delta(0)=2.0\pm0.08\,k_B T_c$ at 3.2\,T and $\Delta(0)=1.73\pm0.08\,k_B T_c$ at 7\,T. Compared to the gap at zero field $\Delta(0)=2.1 k_B T_c$, from tunneling measurements~\cite{Orl.79}, we can see that the energy gap in the low-temperature limit decreases monotonically with the field, indicating  suppression of the order parameter, consistent with strong coupling character of the Nb$_3$Sn superconductor.

	\section{Conclusion}
	In summary, a high-quality Nb$_3$Sn powder sample with cubic structure was investigated by nuclear magnetic resonance. Anisotropy in the Knight shift was found from measurement and analysis of the $^{93}$Nb NMR spectrum. It was also found that under certain temperature and field conditions, the powder grains can be aligned to some extent due to the presence of magnetic anisotropy. The Knight shift and $T_1^{-1}$ in the normal state were measured. The decrease of Knight shift on cooling indicates that core polarization is enhanced at lower temperatures. In the superconducting state, the measurement of the spin-lattice relaxation rate, $T_1^{-1}$, is compared to BCS theory, giving an energy gap $\Delta(0)=2.0\pm0.08\,k_B T_c$ at 3.2\,T and $\Delta(0)=1.73\pm0.08\,k_B T_c$ at 7\,T, consistent with suppression of the order parameter by magnetic field. In addition, quasiparticle coherence effects are observed just below $T_c$ in both 3.2\,T and 7\,T, and absence of a Hebel-Slichter peak.

	\section{Acknowledgement}
	This work was supported by the U.S. Department of Energy, Office of Science, National Quantum Information  Science Research Centers, Superconducting Quantum Materials and Systems Center (SQMS) under contract No. DE-AC02-07CH11359. A portion of this work was performed at the National High Magnetic Field Laboratory, which is supported by National Science Foundation Cooperative Agreement No. DMR-2128556 and the State of Florida.
	
	\bibliography{Nb3Sn_Paper}
\end{document}